\documentclass[runningheads]{svmult}

\usepackage{makeidx}   
\usepackage{graphicx}  
\usepackage{subeqnar}  
\usepackage{multicol}  
\usepackage{physprbb}  
\makeindex             

\newcommand{\Msun}{\mbox{${\cal M}_\odot$}}

\newcommand{\Halpha}{\mbox{H{\footnotesize $\alpha$}}}
\newcommand{\Hbeta}{\mbox{H{\footnotesize $\beta$}}}
\newcommand{\Pab}{\mbox{Pa{\footnotesize $\beta$}}}

\newcommand{\Brg}{\mbox{Br{\footnotesize $\gamma$}}}
\def\lesssim{\mathrel{\hbox{\rlap{\hbox{\lower4pt\hbox{$\sim$}}}\hbox{$<$}}}}
\def\gtrsim{\mathrel{\hbox{\rlap{\hbox{\lower4pt\hbox{$\sim$}}}\hbox{$>$}}}}

\begin{document}

\title*{How to Correct for Dust Absorption in Starbursts}
\toctitle{How to Correct for Dust Absorption in Starbursts}

\author{Gerhardt R. Meurer \and 
Mark Seibert}

\authorrunning{G.R.\ Meurer \&\ M.\ Seibert}

\institute{The Johns Hopkins University, Baltimore MD 21218, USA}

\maketitle

\begin{abstract}
We review new and published results to examine how well the bolometric
flux of starbursts can be recovered from ultraviolet (UV) and optical
observations.  We show that the effective absorption of starbursts can
be substantial, up to $\sim 10$~mag in the far UV, and $\sim 5$~mag in
\Halpha, but apparently not as high as some claims in the literature
(several tens to a thousand mag). The bolometric fluxes of an IUE sample
of starbursts can be recovered to 0.14~dex accuracy using the UV
flux and spectral slope. However, this relationship breaks down for
Ultra Luminous Infrared Galaxies (ULIGs).  The \Halpha\ flux combined
with the Balmer decrement can be used to predict the bolometric flux
to 0.5~dex accuracy for starbursts including most ULIGs.  These
results imply a foreground screen component to the dust distribution.
\end{abstract}

\section{Introduction}

Dust presents one of the biggest obstacles to interpreting observations
of starburst galaxies in the optical and especially the ultraviolet
(UV).  The problem is difficult, because it depends not only on the
amount of dust and its composition, but also the distribution of both
dust and light sources.  Faced with such complexity, the astronomical
community's response includes assuming that star formation remains
mostly unobscured by dust~\cite{madau96} to deriding those who even
consider that UV and optical observations can be corrected for
dust~\cite{wtc92}.

Here we use new and published observations to critically examine the
importance of dust absorption in starburst galaxies, in order to answer
the following: 1.\ Is dust important? 2.\ What is the dust geometry? 3.\
How does dust effect broad band colors and fluxes? 4. What does it do to
emission line fluxes and line ratios?  5.\ \emph{Can we recover the
bolometric flux of a starburst from its UV or optical properties?\/}.
The last question is a proxy for asking whether we can determine the
star formation rate, but avoids distance uncertainties and assumptions
about the lower end of the Initial Mass Function (IMF).

\section{Samples and Tracers}

\begin{figure}[ht]
\begin{center}
\includegraphics[width=0.48\textwidth]{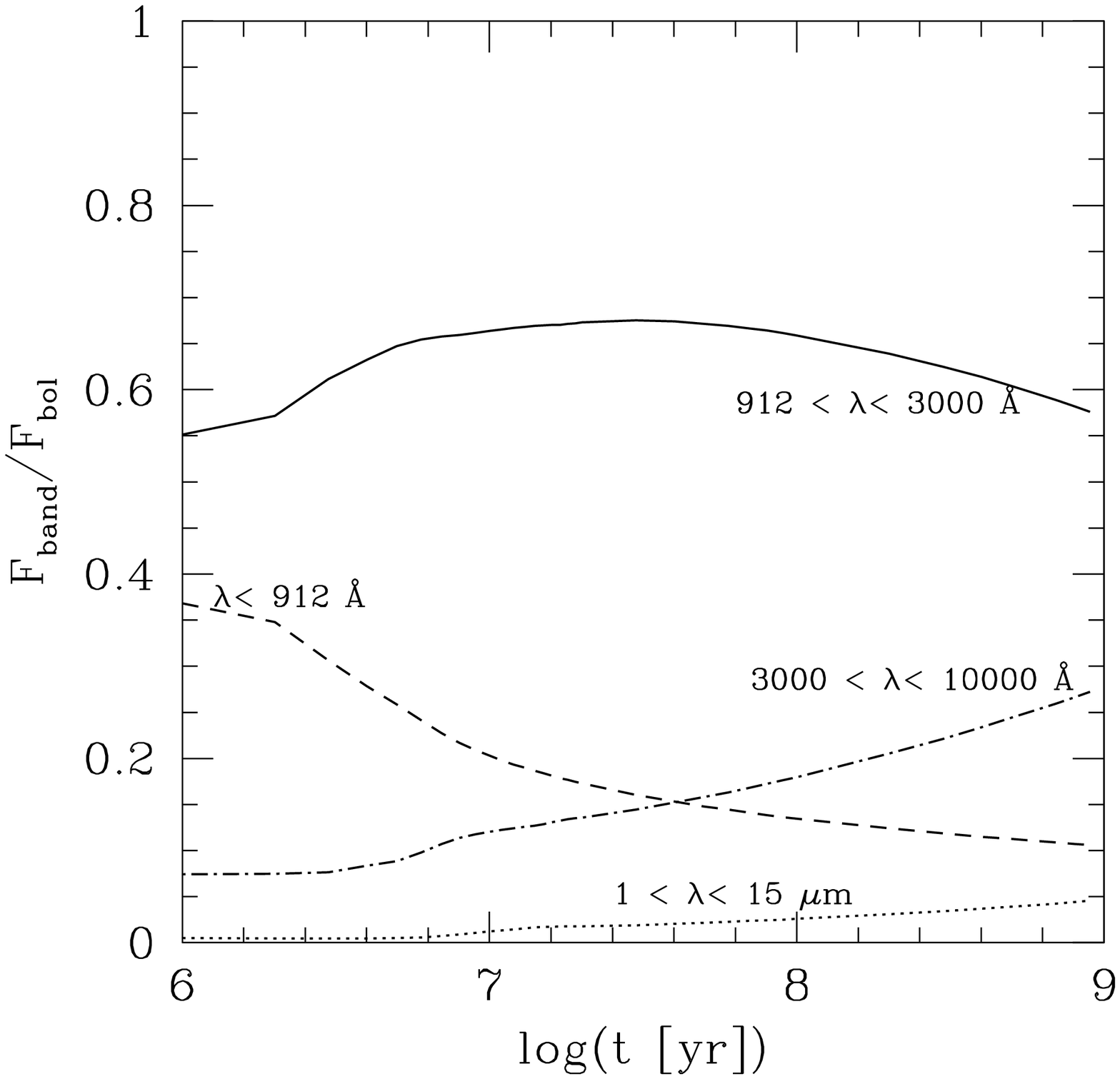}
\includegraphics[width=0.48\textwidth]{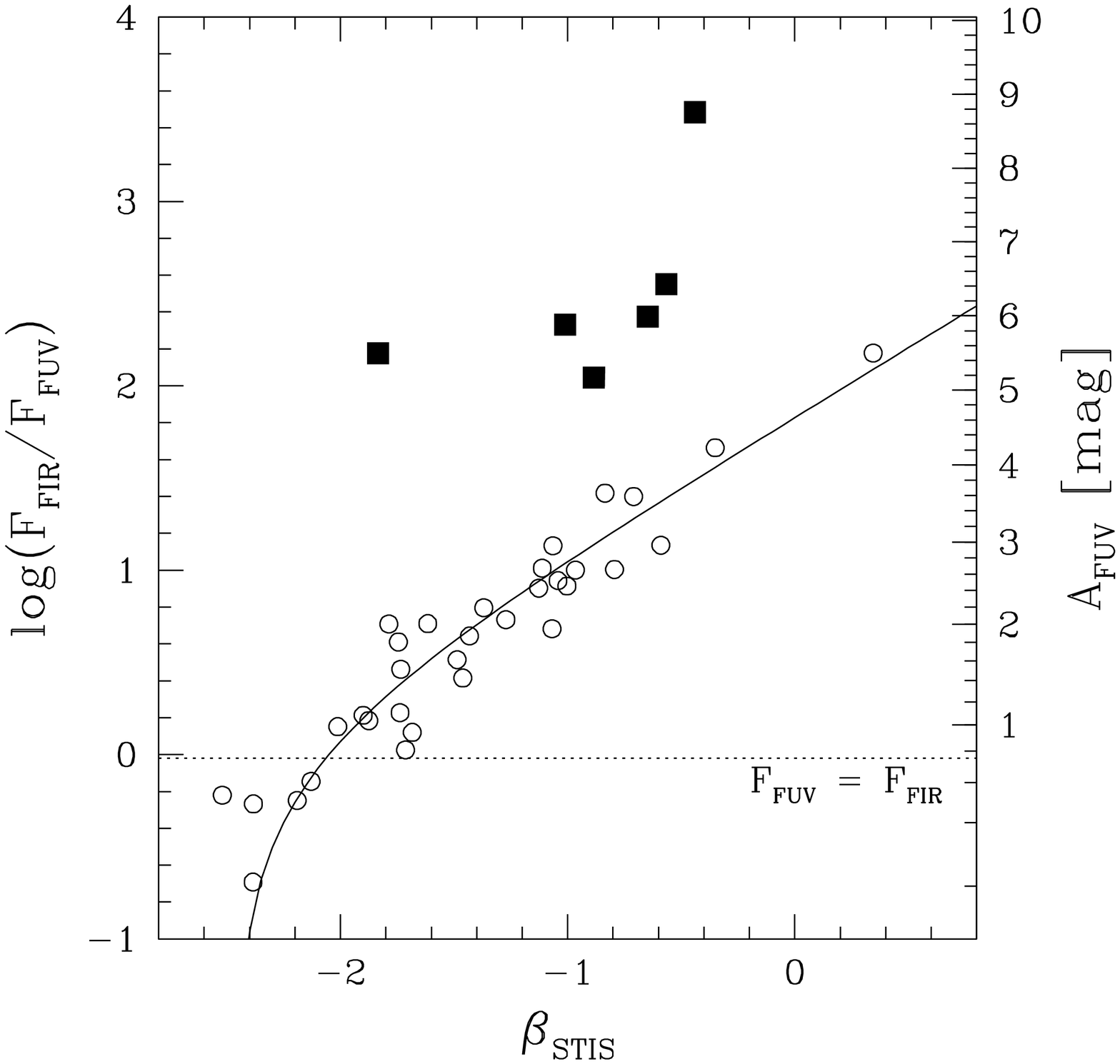}
\end{center}
\caption[]{{\bf a. (Left)}: The fraction of the intrinsic bolometric
luminosity emitted as ionizing radiation (dashed), UV (solid), optical
(dot-dashed), and infrared (dotted) for a constant star formation rate
stellar population with a Salpeter IMF (limits 1--100~\Msun) and solar
metallicity~\cite{starburst99}. {\bf b. (Right)}: Ratio of FIR to far UV
(FUV) flux plotted against spectral slope $\beta_{\rm STIS}$ for IUE
starbursts (open circles) and ULIGs (squares).  Here the FUV flux,
$F_{\rm FUV}$ and $\beta_{\rm STIS}$ are derived from the STIS
bandpasses used for the ULIG observations~\cite{g01}.  The right axis
shows the effective UV absorption.  The solid line shows a linear fit of
$A_{FUV}$ to $\beta_{\rm STIS}$ to the IUE sample. The horizontal line
shows where the bolometric corrected UV and FIR fluxes are equal.}
\label{introfigs}
\end{figure}

We consider two samples of starburst galaxies, those observed in the UV
by the \emph{International Ultraviolet Explorer\/} (IUE), and
starbursts found in the far-infrared (FIR) by the \emph{InfraRed
Astronomical Satellite\/} (IRAS).  The samples are complementary.  The
IUE sample contains a lot of dwarfs as well as starbursts with $L_{\rm
bol}$ as high as $\sim 10^{11.5}~L_\odot$.  IUE starbursts are good
templates for high-redshift Lyman Break Galaxies~\cite{m97,m99}.  The IRAS sample
includes the most luminous starbursts, the Ultra-Luminous Infrared
Galaxies (ULIGs: $L_{\rm bol} \geq 10^{12}~L_\odot$), but very few
dwarfs. ULIGs make up $< 6$\%\ of the FIR background~\cite{sm96}, but may be good
templates for high-redshift sub-mm sources which could contribute
significantly to the star formation rate density at $z \gtrsim
2$~\cite{barger98}. 

We consider two tracers of star formation: the UV continuum, and Balmer
emission lines.  The UV continuum dominates the \emph{intrinsic\/}
(before dust) bolometric output of starbursts (Fig.~\ref{introfigs}a),
and is sensitive to main sequence stars with ${\cal M}_* > 5~\Msun$.
The Balmer lines are among the strongest in the optical and provide a
good measure of the ionizing flux, and hence to stars having ${\cal M}_*
> 20~\Msun$.

\section{The IUE Starburst Sample}

The FIR emission of a starburst represents the total luminosity absorbed
by dust.  For any star forming or young population, the dust heating is
dominated by the UV, hence the FIR/UV flux ratio, or infrared excess
(IRX) can be transformed directly into an ``effective absorption''.
Figure~\ref{introfigs}b plots IRX versus UV spectral slope $\beta$
($f_\lambda \propto \lambda^\beta$) for both samples.  Here $F_{FUV}$ is
a generalized flux $\lambda f_\lambda$ evaluated at rest wavelength
$\lambda = 1515$\AA, where $f_\lambda$ is the flux density.  Almost all
galaxies emit more in the FIR than the UV, hence dust is clearly
important for defining the spectral energy distribution.  A strong
correlation, the IRX-$\beta$ relationship, is apparent for the IUE
sample and readily fit by a linear relationship between effective UV
absorption and $\beta$~\cite{m95,m99}.

\begin{figure}[ht]
\begin{center}
\includegraphics[width=0.48\textwidth]{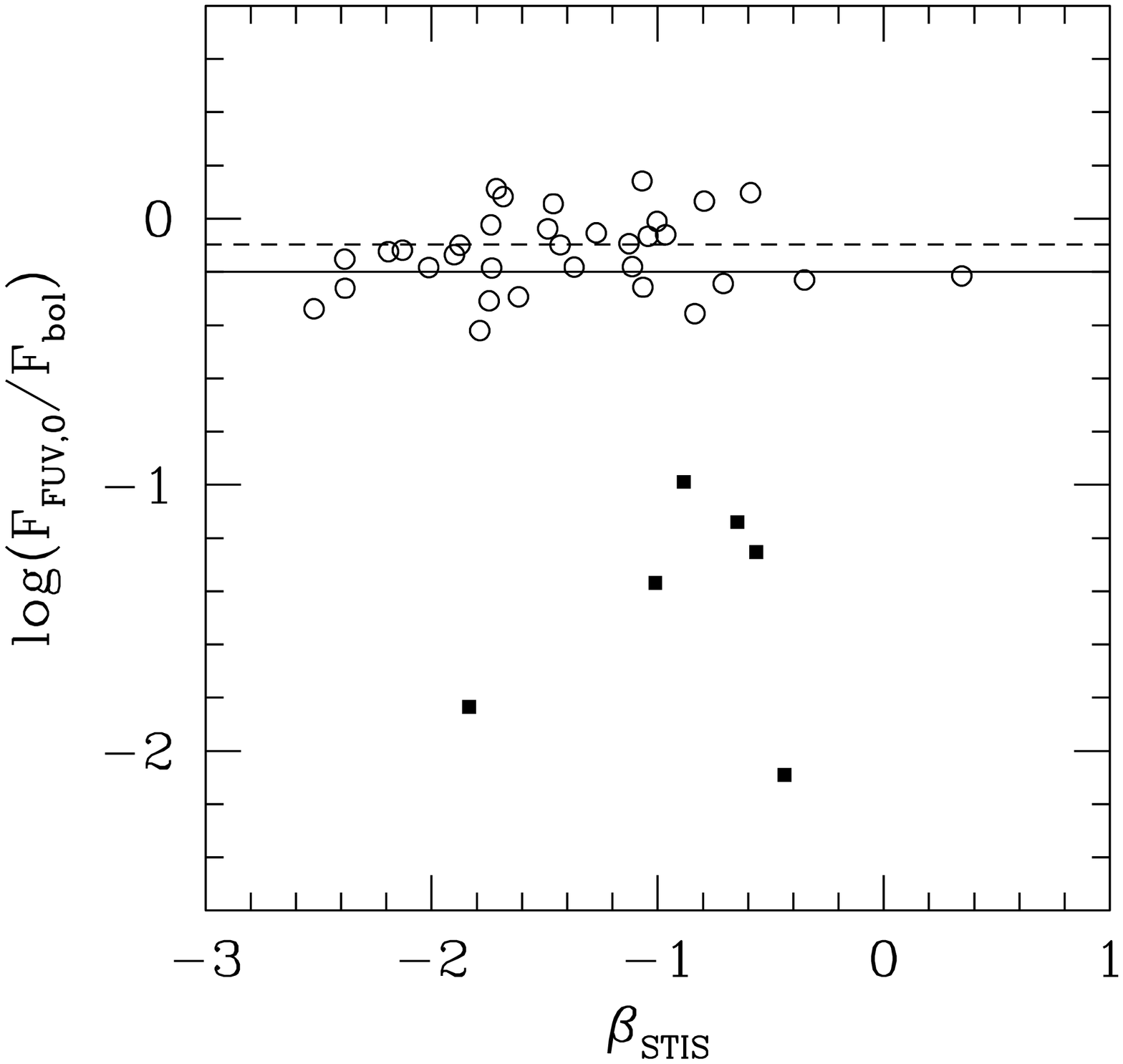}
\includegraphics[width=0.48\textwidth]{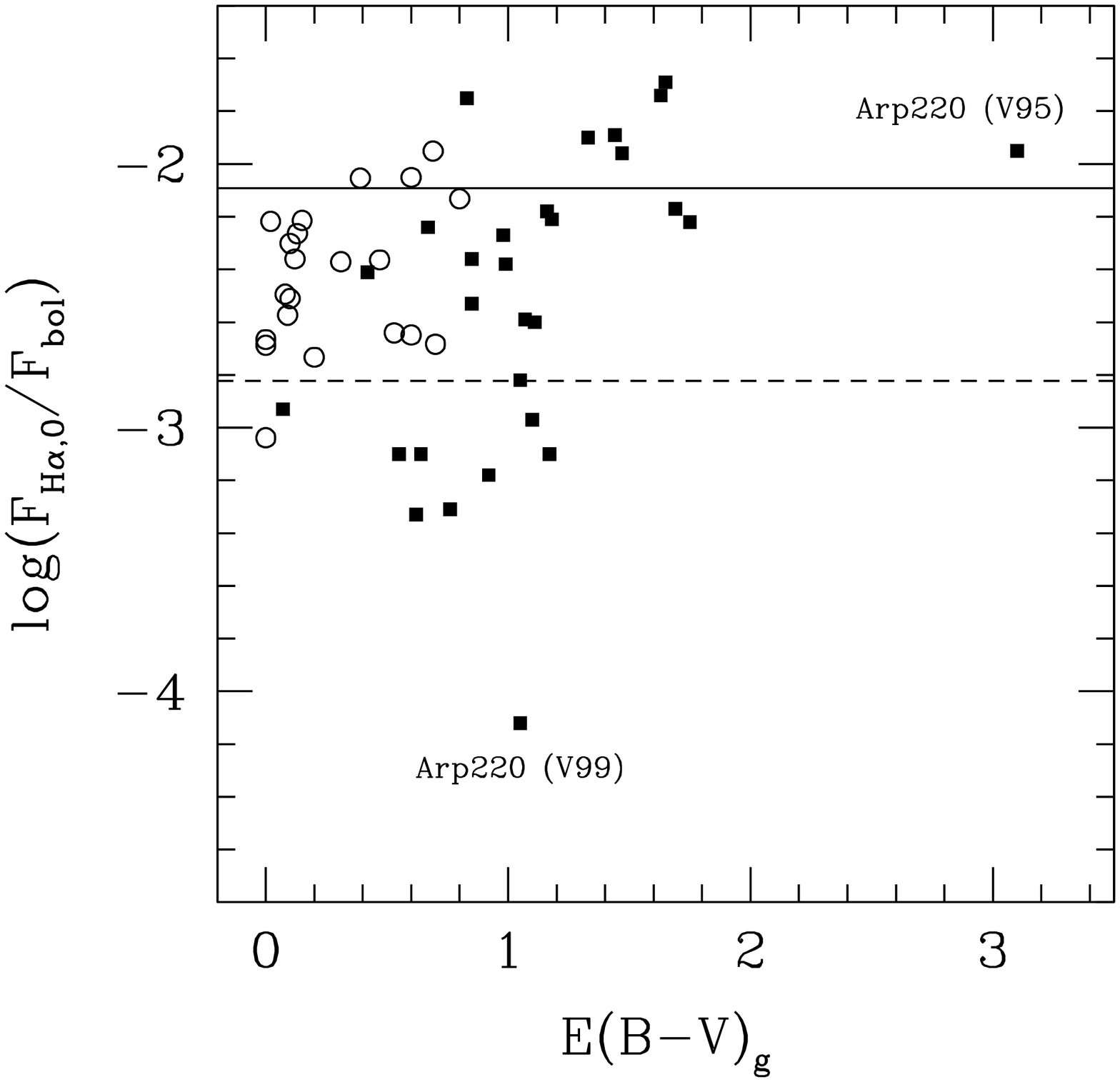}
\end{center}
\caption[]{Ratio of recovered flux (observed flux corrected for dust
absorption as deduced from reddening) to bolometric flux, $F_{\rm bol}$,
plotted against reddening indicator.  Symbols are as in
Fig.~\ref{introfigs}b. For the IUE sample $F_{\rm bol}$ is a weighted
sum of the observed UV and FIR fluxes, for the IRAS sample only the FIR
flux is used.  In {\bf Fig~\ref{bothsamp}a (left)}, the numerator is the
UV flux and the ratio is compared to the UV spectral slope $\beta$.  In
{\bf Fig.~\ref{bothsamp}b (right)}, the numerator is the \Halpha\ flux
and the ratio is compared to the reddening $E(B-V)_g$ determined by the
\Halpha/\Hbeta\ decrement.  The horizontal lines show model predictions
from Starburst99~\cite{starburst99} for a stellar population forming at
a constant star formation rate for 100~Myr and having a Salpeter IMF
with a lower mass limit of 1~\Msun\ and upper mass limits of 100~\Msun\
(solid line) and 30~\Msun\ (dashed line).}
\label{bothsamp}
\end{figure}

Figure~\ref{bothsamp}a plots the absorption corrected UV flux,
$F_{FUV,0}$ to $F_{\rm bol}$ ratio versus $\beta$.  The absorption comes
from the fit plotted in Fig.~\ref{introfigs}b.  The mean logarithmic
ratio of the IUE sample is $\langle \log(F_{FUV,0}/F_{\rm bol}) \rangle
= -0.13$ with a scatter of 0.14~dex and no residual correlation with
$\beta$.  For the IUE starbursts, the FUV flux and $\beta$ are
sufficient to recover the $F_{\rm bol}$ to 40\% accuracy.
For this sample Balmer fluxes from IUE aperture matched
spectra~\cite{mck95,skc95} are available.  Figure~\ref{bothsamp}b
compares the ratio of dust corrected \Halpha\ flux $F_{\rm H\alpha,0}$
and $F_{\rm bol}$ with $E(B-V)_g$, the intrinsic reddening of the
ionized gas.  Again, there is no correlation between the two quantities.
Note that the dust correction, $A_{\rm H\alpha} = 2.5 E(B-V)_g$,
assumes a standard Galactic extinction law.  The mean logarithmic ratio
is $\langle \log(F_{\rm H\alpha,0}/F_{\rm bol}) \rangle = -2.43$ with a
dispersion of 0.28 dex. The Balmer fluxes can recover the $F_{\rm bol}$
to better than a factor of 2 in this sample.  Calzetti et al.~\cite{c96}
find that $E(B-V)_g$ measured from IR \Pab\ and \Brg\ to Balmer line
ratios agrees well with that measured from only the Balmer lines.

\section{The FIR Selected Starburst Sample}

Very few UV observations of ULIGs exist, perhaps because they were
expected to be so dusty as to be invisible in the UV.  However recent
observations of ULIGs from the ground at $\lambda_c \sim
3400$~\AA~\cite{ss00}, and with HST at $\lambda_c \sim 2300,
1400$~\AA~\cite{tks99} show that ULIGs do emit a small fraction of their
bolometric luminosity in the UV.

We obtained \emph{Space Telescope Imaging Spectrograph\/} UV images of
seven galaxies with $L_{\rm bol} \geq 10^{11.6}~L_\odot$~\cite{g01}.  
In all cases the galaxies were detected
in both the far UV (rest $\lambda_c \sim 1515$~\AA) and the near UV (rest
$\lambda_c \sim 2440$~\AA) with some UV emission detected within a kpc of
the nuclei~\cite{sco00}.  However, in most cases the UV peak does not
coincide well on the few hundred parcecs scale with the near IR emission.
Figures~\ref{introfigs}b and~\ref{bothsamp}a show that the IRX-$\beta$
correlation underpredicts the bolometric flux of ULIGs by a factor
ranging from $\sim$~7 to~90.

The situation is more optimistic with Balmer lines as shown in
Fig.~\ref{bothsamp}b which includes data on 28 IRAS galaxies with
$L_{\rm bol} > 10^{11.6}~L_\odot$.  
The \Halpha\ fluxes were derived from narrow
band images~\cite{ahm90}, while spectra from a variety of published
sources~\cite{v95,v99,wu98b} were used to remove [{\sc Nii}]
contamination and measure \mbox{$E(B-V)_g$}.  After correcting
for absorption, ULIGs have similar $\langle \log(F_{\rm
H\alpha,0},F_{\rm bol})\rangle = -2.48$ to the IUE starbursts, 
with a somewhat larger dispersion: 0.51 dex.

One caveat is that these points represent total \Halpha\ fluxes
corrected with \emph{nuclear} ($R < 1$~kpc) line ratios.  Gradients in
$E(B-V)_g$ may mean that we overestimated $F_{\rm H\alpha,0}$.  Large
spatial variations certainly exist in some galaxies as shown by the two
data point for Arp220 in Fig.~\ref{bothsamp}b.  However, on average the
gradients are shallow with typically $\Delta E(B-V)_g \approx 0.4$ mag
out to $R = 8$ kpc (where the contribution to the total \Halpha\ flux is
small) compared to $\langle E(B-V)_g \rangle \approx 1.1$ mag in the
center~\cite{kvs98}.  Clearly, spectroscopy over the entire face of ULIGs
is required to properly determine $F_{\rm H\alpha,0}$.  There is
precious little of this available in the literature.  While we can not
yet rule out a fortuitous coincidence, Fig.~\ref{bothsamp}b indicates
that integrated Balmer line fluxes can be used to predict the bolometric
flux of starbursts to a factor of about three accuracy, even in most
ULIGs.


\section{Discussion}

UV color and/or Balmer line ratios can be used to crudely
estimate the $F_{\rm bol}$ of starbursts, even ULIGs.  When looking at
large samples an accuracy of $\sim$ 0.5~dex should be sufficient for
measuring integrated star formation rate densities. The effective
absorption implied by Figs.~\ref{introfigs}b~\&~\ref{bothsamp}b is
$\lesssim 10$~mag in the far UV, and $\lesssim 5$~mag in \Halpha.  Five
to ten magnitudes of dust absorption is large (factor of~$10^2$ to~$10^4$ in
attenuation), but not overwhelming.  These results contradict the
notion that star formation is essentially buried behind unmeasurabley
large absorption in the UV and optical.  Why is this?

First of all, our results do not rule out some very buried star
formation.  Perhaps the completely buried phase is of short duration,
before stars migrate from their birth site or the surrounding dust and
gas is cleared away by the supernovae~\cite{kj00}.  This scenario may
also explain the higher dust column density affecting emission lines
compared to continuum radiation~\cite{cf00}.  However, we have not found
cases where all the star formation is buried behind several tens of
magnitudes of absorption in both the UV and \Halpha.  Some appropriately
reddened emission usually gets out.  Secondly, some claims for missing
star formation are very model dependent.  For example
Poggianti~\cite{pog00} mentioned a short fall of a factor of three in
\Halpha\ derived star formation rates compared to FIR estimates.
However this is relative to a stellar population model of constant star
formation rate with an assumed IMF upper mass limit ${\cal M}_u =
100$~\Msun.  A deficit of 0.5 dex compared to this model is completely
consistent with our results (Fig.~\ref{bothsamp}b) which
suggests ${\cal M}_u \approx 50$~\Msun\ may be more appropriate.
Aperture effects may be behind other claims of high extinction.  For
example Sturm et al.~\cite{sturm96} use flux ratios of weak near to mid
IR recombination and fine structure lines to infer a $V$
band dust absorption of 30--80 mag for Arp220.  However, the aperture size they use
increases with wavelength, which can induce a spuriously large
absorption since this source fills these apertures in \Halpha~\cite{ahm90}.

The correlation of effective absorption with $\beta$ (IUE sample) and
Balmer decrement (both samples) strongly suggests a foreground screen
dust geometry~\cite{c94,c96}.  While there is some hostility to such a
model (e.g.\ ~\cite{wtc92}), we have yet to see these correlations well
modeled without a screen contribution.  However, this screen is not
likely to be a thin uniform sheet encompassing all star formation
tracers, otherwise the ULIGs would fall on the same IRX-$\beta$
relationship as the IUE sample.  The Charlot and Fall model~\cite{cf00}
is a hybrid containing both foreground screen dust shells and mixed gas
and dust. It works well for the IUE galaxies and perhaps can be adapted
to fit the IRAS sample as well.

It should be no surprise that a foreground screen component is required,
since a starburst can naturally produce such a screen.  Its stellar
winds and supernovae will evacuate a cavity around the starburst and
power a galactic wind~\cite{ham90}.  Most of the dust opacity will arise
in the walls of this cavity.  Any molecular clouds that wander into the
cavity will be compressed by the high pressure within the starburst and
hence have a low covering factor.  Direct evidence for this scenario is
given by Heckman et al.~\cite{hlsa00} who show that the metal content in
the wind is directly related to the reddening.  In particular, the depth
of the blue shifted Na{\sc i} absorption line in starbursts correlates
well with both the optical continuum color and the $\Halpha/\Hbeta$
ratio.

\section{Conclusions}

We conclude by answering the questions we posed at the start: 1.\ Yes,
dust is important in most starbursts. 2.\ The dust geometry 
includes a strong foreground screen contribution, probably arising in a
galactic wind.  3.\ Dust reddens the UV colors as the flux is diminished
in the IUE starbursts, but this relationship breaks down for ULIGs.  4.\
Optical emission line flux ratios redden with increasing dust absorption
for all types of starbursts.  5.\ The bolometric flux of starbursts can
be recovered from their UV color (except ULIGs) or more crudely, from
Balmer line flux ratios (all starbursts).  These results bode well for
estimating the star formation rate density locally and out to high
redshift from UV and optical surveys.

\smallskip

{\bf Acknowledgements}. We thank the conference organizers for the
opportunity to research this topic.  This contribution is a progression
from what was shown at the meeting (including more points in
Fig.~\ref{bothsamp}b) where GRM lead the discussion on this topic.  We
thank Kurt Adelberger, Tim Heckman, Bianca Poggianti and Sylvain Veilleux for stimulating
conversations that helped refine our case.  We also thank our
collaborators on the STIS ULIG project, Jeff Goldader, Tim Heckman,
Daniela Calzetti, Dave Sanders, and Chuck Steidel.

\end{document}